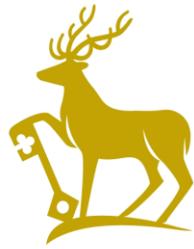

# UNIVERSITY OF SURREY

Towards automated identification of changes in
laboratory measurement of renal function:
implications for longitudinal research and
observing trends in glomerular filtration rate (GFR)

Norman Poh   Andrew McGovern   Simon de Lusignan

SEPTEMBER 2014

TR-14-03

# Department of Computing
# Technical Report

# Towards automated identification of changes in laboratory measurement of renal function: implications for longitudinal research and observing trends in glomerular filtration rate (GFR)

## Author List

Dr Norman Poh[a]
Lecturer in Database and Knowledge Discovery
NP      normanpoh@ieee.org

Dr Andrew McGovern[b]
Researcher in Primary Care
AMcG    andy@mcgov.co.uk

Professor Simon de Lusignan[b]*
Professor of Primary Care and Clinical Informatics
SdeL    s.lusignan@surrey.ac.uk

*Author for correspondence

[a] Department of Computing
University of Surrey,
GUILDFORD
Surrey              GU2 7XH         UK

[b] Department of Healthcare Management and Policy
University of Surrey,
GUILDFORD
Surrey              GU2 7XH         UK

# Abstract

**Introduction:** Kidney function is reported using estimates of glomerular filtration rate (eGFR). However, eGFR values are recorded without reference to the creatinine (SCr) assays used to derive them, and newer assays were introduced at different time points across laboratories in UK. These changes may cause systematic bias in eGFR reported in routinely collected data; even though laboratory reported eGFR values have a correction factor applied.

**Design:** An algorithm to detect changes in SCr which affect eGFR calculation method by comparing the mapping of SCr values on to eGFR values across a time-series of paired eGFR and SCr measurements.

**Setting:** Routinely collected primary care data from 20,000 people with the richest renal function data from the Quality Improvement in Chronic Kidney Disease (QICKD) trial.

**Results:** The algorithm identified a change in eGFR calculation method in 80 (63%) of the 127 included practices. This change was identified in 4,736 (23.7%) patient time series analysed. This change in calibration method was found to cause a significant step change in reported eGFR values producing a systematic bias. eGFR values could not be recalibrated by applying the Modification of Diet in Renal Disease (MDRD) equation to the laboratory reported SCr values.

**Conclusions:** This algorithm can identify laboratory changes in eGFR calculation methods and changes in SCr assay. Failure to account for these changes may misconstrue renal function changes over time. Researchers using routine eGFR data should account for these effects.

Word count: 235

# Introduction

Chronic kidney disease (CKD) is a significant public health problem and is becoming more common with an aging population and increasing disease burden from diabetes (1, 2). There is a complex relationship between CKD, diabetes, and hypertension resulting in increased risk of mortality and cardiovascular disease in people with these commonly co-morbid conditions (3). Recent estimates of the prevalence of CKD in the UK are around 7.3-8.5% (4, 5). This is associated with substantial financial burden: In 2009-2010 the cost of CKD to the English NHS was estimated at £1.44-£1.45 billion, approximately 1.3% of the total NHS spending during this period (6). Over half of this was spent on renal replacement therapy for people with end stage renal disease (ESDR), which accounts for only 2% of the CKD population (6). Early identification, appropriate referral, and intervention in CKD are therefore critically important.

Estimation of renal function has been routine in clinical practice since the publication of the Cockcroft-Gault equation for estimating creatinine clearance in 1976 (7). Categorisation of CKD and clinical decisions are currently based on estimated glomerular filtration rate (eGFR) (8, 9) although the Cockcroft-Gault equation is still widely used to calculate drug dosing (10, 11). eGFR can be calculated from serum creatinine (SCr) measurements using the Modification of Diet in Renal Disease (MDRD) equation first published in 1999 (12) and later simplified in 2003 (13). However, the MDRD equation underestimates GFR in people with mild renal impairment (14, 15) and in some subgroups such as kidney donors and people with diabetes (16, 17). More recently the Chronic Kidney Disease Epidemiology Collaboration (CKD-EPI) equation was developed to tackle these limitations and has been demonstrated to have improved performance in mild renal impairment and across patient subgroups (18-21).

Continuously changing methods for calculating renal function present a problem for both clinicians and researchers. Changing methods of eGFR calculation affect trends in renal function over a period of years. This problem is further compounded by differing creatinine assays between laboratories (22, 23). The UK National External Quality Assessment Scheme (UKNEQAS) recommend each clinical laboratory calculate eGFR from an isotope dilution mass spectrometry (IDMS) creatinine assay; a standardisation program which was initiated in 2007 (24, 25) with different laboratories achieving standardisation at different times. In clinical practice, laboratory calculated eGFR values are reported without reference to the equations or creatinine assays that were used to derive them.  Records will also contain eGFR results derived in practice possibly using one of the many online calculators.  This practice was very common when primary care first became aware of CKD (26).

Before the introduction of the MDRD equation there was no reported eGFR. Following the introduction of this equation laboratories were reporting eGFR calculated from various inhomogeneous creatinine assays. These were then phased out to be replaced by a standardised method based on the IDMS assay (Figure 1).

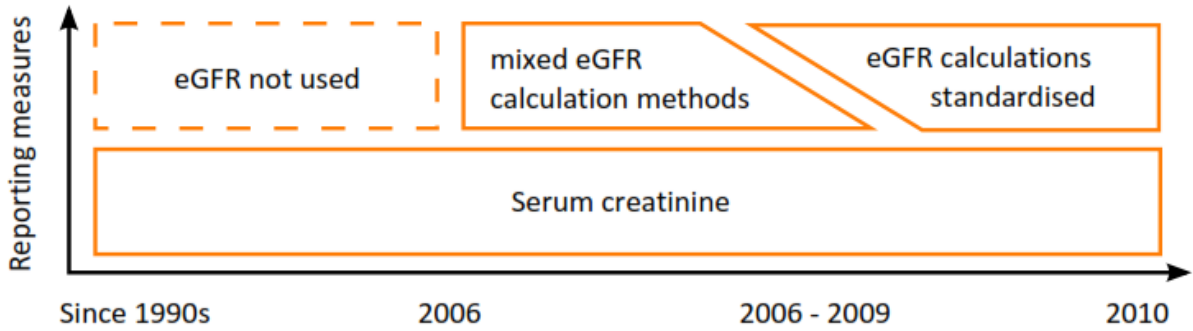

Figure 1: Changes in reporting of serum creatinine and eGFR data over time. Since the 1990's, various versions of serum creatinine measurements have been systematically recorded. Since 2006, eGFR has been used. Between 2006 and 2009, creatinine assays have been standardised and correction factors applied before using the MDRD have changed with the assay method.

Current clinical guidelines recommend early referral of patients with declining GFR to specialist services. Whilst standardisation of current eGFR values has been achieved there is a need for retrospective calibration both in the research and clinical settings to allow accurate monitoring of renal function trends. The previous existence of these heterogeneous assays prevents accurate assessment of trends in retrospective data because these assays are incompatible with each other. Should the CKD-EPI equation be widely adopted in primary care this will be of renewed importance (27). Here we devise a method for identifying changes in eGFR calculation method (which includes correction factors for the creatinine assay used) in routinely collected data. Identification of these changes in eGFR calculation method is the first step towards backwards calibration of the entire eGFR time-series for a given patient – i.e. making all of the patient's eGFR measurements compatible. Without such a method trends in renal function are misleading. Our algorithm identifies the date of change from one method eGFR calculation method to the next for each patient and primary care practice.

## Method

The UK electronic patient record is currently coded using the Read coding system. This enables coding of the eGFR equation used, although eGFR can be coded with no reference to the equation used. We investigated the range of codes available to record eGFR to explore if there was scope to improve the provenance of these data.

We devised an algorithm which is able to identify the changes in calculation method of eGFR for any time series of eGFR and SCr measurements for a given patient. This method requires a time series of paired eGFR and SCr values for each patient i.e. SCr and eGFR values measured simultaneously. We shall call this paired measurement time series the renal function time series of the patient.

### The eGFR method change finding algorithm

Each laboratory uses a function ($M_{lab}$) to convert SCr measurements ($c$) into eGFR values ($g_{lab}$). Thus:

$$g_{lab} = M_{lab}(c, CF) \qquad (1)$$

where *CF* represents patient-specific adjustment based on patient characteristics (age, gender, and ethnicity). To identify changes in $M_{lab}$ we defined a self-calculated eGFR ($g_{self}$) generated using the MDRD equation ($M_{MDRD}$) and using patient characteristics (*CF*) taken from the patient record:

$$g_{self} = M_{MDRD}(c, CF) \quad (2)$$

We then defined a mapping function ($M_u$) which maps the self-calculated eGFR onto the lab calculated eGFR (Figure 2). The lab calculated eGFR can therefore also be written as:

$$g_{lab} = M_u(g_{self}) \quad (3)$$

The mapping function $M_u$ will vary with the laboratory eGFR calculation function, $M_{lab}$. It is therefore possible to determine the number of laboratory $M_{lab}$ functions by determining how many $M_u$ functions are required to map the entire renal function time series. We identified that $M_u$ is a linear function in the logarithmic domain of its argument, $g_{lab}$ and $g_{self}$. Let us define the mapping function from $\log(g_{lab})$ to $\log(g_{self})$ be $M_u'$. This function must be linear[1], taking the form of:

$$M_u'(\log(g_{lab})) = w_1 \log(g_{self}) + w_0 \quad (4)$$

where the parameters $w_1$ and $w_0$ can be found by the method of least square regression in the case of a single assay method and a mixture of regression (28) in the case of multiple assay methods. The lab calculated eGFR can therefore be obtained by:

$$g_{lab} = M_u(g_{self}) = \exp(w_1 \log(g_{self}) + w_0) \quad (5)$$

Performing this mapping from $g_{self}$ onto $g_{lab}$ for a series of two or more measurements enables $w_1$ and $w_0$ to be calculated. The number of $w_1$ and $w_0$ value pairs required to map a renal function time series for a given patient is the number of laboratory eGFR calculation functions used.

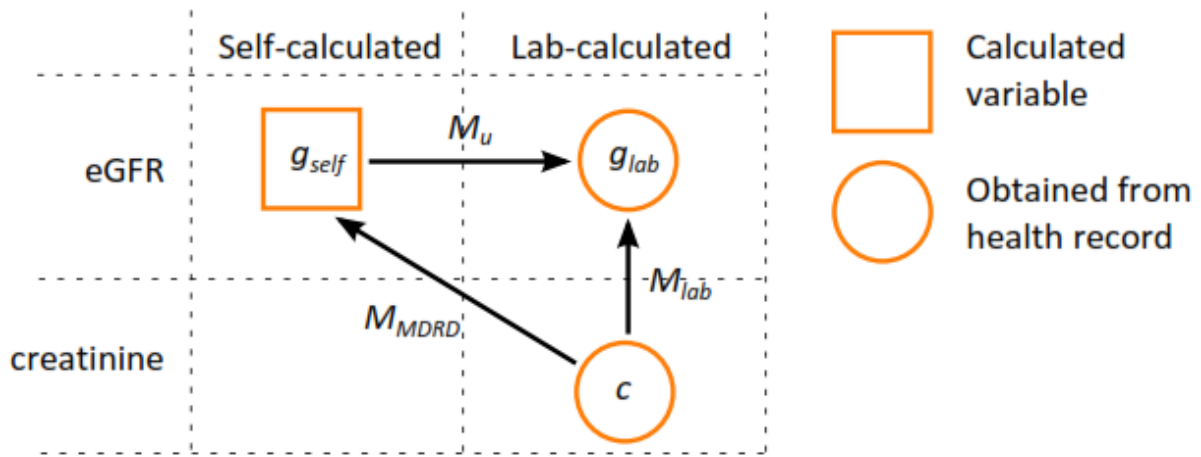

Figure 2: The relationship between different variables. $g_{self}$ refers to self-calculated eGFR using the MDRD; $g_{lab}$, the lab calculated eGFR; *c*, serum creatinine; *M*, different mapping functions. Both the $M_{MDRD}$ and $M_{lab}$ also use ethnicity, age, and gender of the patient (not shown here).

---

[1] The proof is omitted but can be easily deduced from MDRD equations in the log domain.

## Subjects and setting

We generated and tested the algorithm using anonymised patient records collected from 127 primary care practices across England; a total of nearly a million patient records (n=951,764). This data was obtained for the Quality Intervention in Chronic Kidney Disease (QICKD) trial (clinical trials registration: ISRCTN56023731) (4). These primary care samples comprise a nationally representative sample of urban, sub-urban and rural practices in localities within London, Surrey, Sussex, Leicester, Birmingham and Cambridge. The complete protocol used for sampling and data collection from these practices for the QICKD trial have been previously described (29). In brief, routine clinical records were collected between June 2008 and December 2010. All practices had a final data collection in December 2010. All patients registered with the included practices at the time of the first the sampling period (June 2008) were included in the data sample. Complete historical records were obtained for all these patients for a number of clinical variables including data relating to renal function. All data was anonymised at the point of data extraction. Data from each practice was labelled with an anonymised practice ID number.

To analyse the usage of eGFR codes we counted the total number of eGFR codes used in the primary care records of all 951,764 people included in the QICKD database.

To test our eGFR calculation change finding algorithm we selected 20,000 patients with the most complete renal function time series in terms of the number of paired laboratory eGFR and SCr.

From the initial patient set, we excluded lab reported values of eGFR readings exactly equal to 60 or 90 ml/min because these values correspond to the capped thresholds chosen by certain laboratories. For example an eGFR value of 93 ml/min, in some cases, are recorded as 60 ml/min by laboratories which adopt the threshold of 60 ml/min or 90 ml/min for others which adopt the higher threshold.

We also used an anonymised practice identification number to group patients by practice. We used this to calculate a time interval between which each practice changed its' eGFR calculation method. As all practices sent laboratory samples to a single laboratory this change will affect all eGFR measurements reported by that practice. The latest identified measurement of the first calculation method and the earliest identified measurement of the second method in each practice were used to define the interval in which the change in method occurred.

## Ethical considerations

No patient identifiable data was used in the analysis presented here. All data was anonymised at the point of data extraction. The original QICKD study was approved by the Oxford Research Ethics Committee (Committee C). This ethics approval included authorisation for secondary analysis of the QICKD dataset.

## Results

A total of 1,309,337 unique eGFR measurements were identified in the QICKD database. The majority (98.7%) of eGFR values were recorded using an equation specific code (Table 1). No codes were identified in the Read code system which enable recording of creatinine assay method.

| Read Code | Code description | Number recorded (%) |
|---|---|---|
| 451F. | Glomerular filtration rate | 16,317 (1.2%) |
| 451E. | Glomerular filtration rate calculated by abbreviated Modification of Diet in Renal Disease Study Group calculation | 1,292,572 (98.7%) |
| 451G. | Glomerular filtration rate calculated by abbreviated Modification of Diet in Renal Disease Study Group calculation adjusted for African American origin | 448 (0.04%) |
| 451K. | Estimated glomerular filtration rate using Chronic Kidney Disease Epidemiology Collaboration (CKD-EPI) formula | 0 (0.0%) |

Table 1. The abundance of 5 byte version 2 Read codes for recording eGFR in the primary care records of 951,764 people. Note the CKD-EPI Read code was not available at the time these data were recorded. Total

The 951,764 patient records obtained from QICKD trial database were available for analysis. For this study, we used the top 20,000 patients who have the renal function time series with the highest number of paired eGFR and SCr values. These 20,000 people included had a median age of 74 (interquartile range; IQR 64 to 81). 10,931 (54.7%) people were female. The median number of SCr measurements per person was 22 (IQR 19 to 28) and the median number of eGFR measurements 16 (IQR 13 to 20). 13,563 (67.8%) people had five or more SCr and lab calculated eGFR values recorded simultaneously.

4,736 (23.7%) people had two distinct detectable methods of calculation of eGFR from SCr. These methods always occurred sequentially with laboratories converting from one method to the other. We did not identify any patients with more than two methods of calculating eGFR.

By grouping patients by their anonymised practice ID we identified the range of dates between which the change in eGFR calculation method occurred for each practice (Figure 3). We identified a change in method in 80 (63%) of 127 included practices.

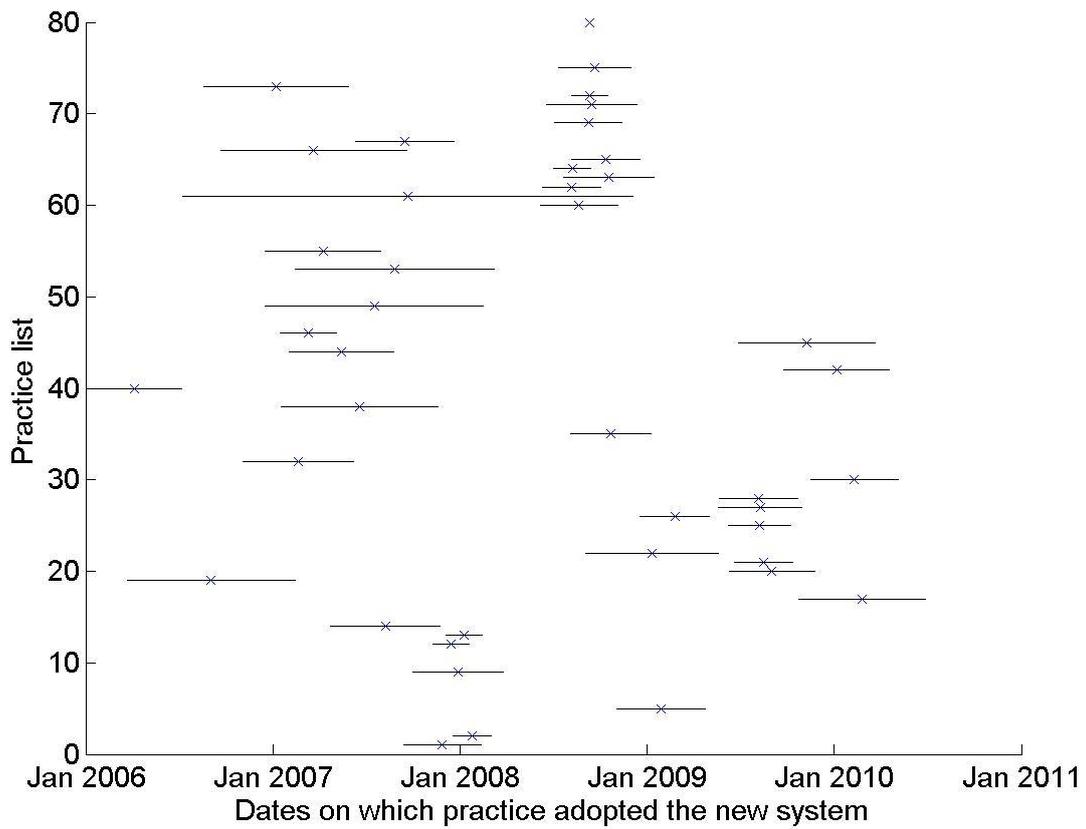

Figure 3: The estimated dates of eGFR calculation method change for each practice where a change was detected in one or more patients within the practice. The range of uncertainty is shown for each practice using a horizontal line.

eGFR time series from patients who had stable renal function both before and after the change in laboratory eGFR calculation method demonstrate a substantial step change at the time of the change in method (Figure 4). Using the MDRD to calculate eGFR from the reported SCr measurements similarly demonstrates a discontinuity (Figure 5). Almost all of the step changes were an improvement in renal function. All of these changes occurred exactly at the time of change in eGFR calculation method. Both of these factors make this observation highly unlikely to be due to an actual physiological change in these patients.

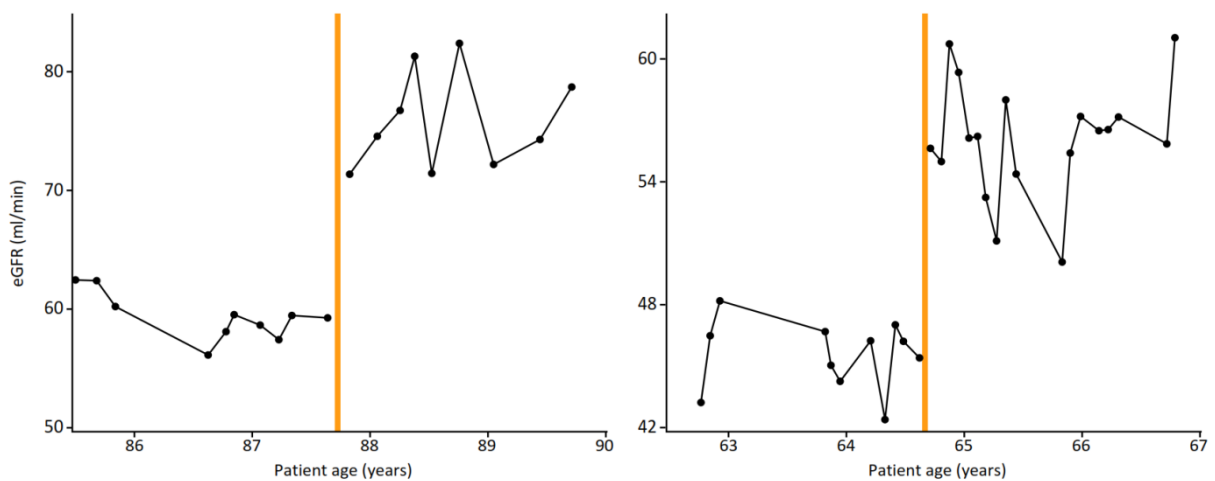

Figure 4: Two examples of selected patients with stable renal function. The change in eGFR calculation method (vertical line) can be seen to coincide with a step discontinuity in eGFR values.

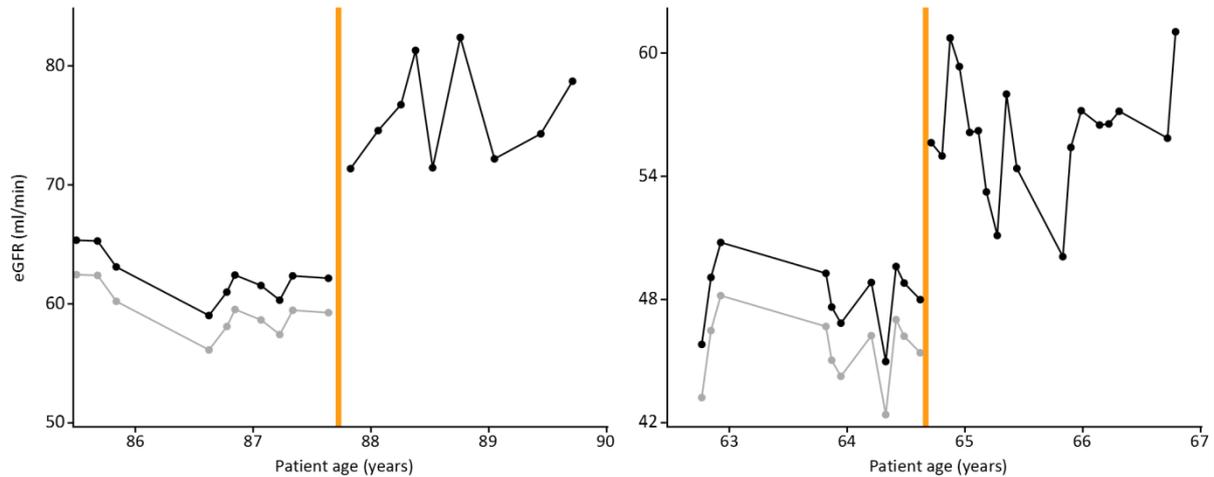

Figure 5: Attempted back calibration by applying the MDRD to the lab reported creatinine values. Recalibration reduces the apparent step discontinuity in these patients compared to the lab reported eGFR values (grey) but a considerable discrepancy is still evident. The patients shown are the same as in Figure 4.

## Discussion

### Principal findings

Our algorithm is able to detect changes in eGFR calculation method from a time-series of eGFR and SCr measurements for any given patient. This change was detected in 23.7% of the people with the highest number of recorded renal function measurements. From these data we identified a change in eGFR calculation method in 63% of the included practices which occurred between January 2006 and December 2010. There were no practices with more than two eGFR calculation methods identified.

Changes in eGFR calculation method create a substantial spurious step change in patient's renal function at the time of change in method. Recalculating eGFR from SCr measurements does not remove this spurious step change. This suggests that the change in eGFR calculation method is also associated with a change in creatinine assay. This is consistent with laboratories responding to the UKNEQAS recommendations to standardise to use the IDMS creatinine assay (25).

The type of equation used to calculate eGFR was generally well recorded but there is currently no way of recording the creatinine method using the Read code system.

### Implications of the findings

Both clinical decisions, such as when to refer to specialist services (8, 9), are often based on renal function trends. Furthermore, there is an increasing amount of research utilising routinely collected data. We have demonstrated that over a third of renal function time series are subject to spurious step changes in renal function as a result of changes in laboratory eGFR calculation methods and creatinine assay changes. If these artefactual changes in renal function are not taken into account in research or clinical decisions utilising longitudinal data of this type there is substantial potential for systematic error.

If information on the type of equation and creatinine assay was required by the eGFR coding structure these artefactual changes would be easy to identify and correct. However whilst the current primary care coding system in the UK (Read codes) does allow this data to recorded this is rarely used. Future recording of eGFR should make use of such features to maximise the clinical and research utility of eGFR measurement and prevent spurious data from impacting on patient care.

### Comparison with the literature
To our best knowledge there have been no previous attempts to detect the time series artefacts we report using large scale population data. The high level of background noise in eGFR measurements (30) mean that these artefacts are not easily identified when observing the data using standard methods.

The importance of coding the context of blood glucose measurements has been previously noted although this call to improve standards has gone unheeded (31). Recording the context of eGFR measurements presents a similar problem. To effectively tackle this issue may require a change in the coding structure of existing coding systems.

### Limitations of the method
The algorithm requires a minimum of two paired eGFR and SCr measurements before a change in eGFR calculation method and two after to correctly identify the change. This limits the population to which the method can be applied. However in dataset where additional information is known, such as the hospital or primary care centre where the test was performed, data from a few patients with a complete renal function time series can be used to predict eGFR calculation method changes in the rest of the population.

Additionally, this method cannot be applied to people with normal renal function as their exact eGFR values are not reported (reported as either >60 ml/min or >90 ml/min). In practices where a change in eGFR calculation method has been detected this change can be assumed affect all members of that practice and could be used to calibrate these data. Furthermore, the exact value of eGFR for this population is of less importance given the poor reliability of eGFR in people with good renal function.

The data used here is now over four years old limiting the clinical utility of this method. However this method is still important for research using longitudinal data. Additionally, as laboratories change from using the MDRD equation to the CKD-EPI equation there will be another period of laboratory eGFR calculation changes. Detection and elimination of the artefacts generated by this new change will be important both clinically and for future epidemiological research with a focus on renal function.

### Call for further research
Our method is the first step to generating a back calibration algorithm which can correct for the different eGFR calculation methods used by different laboratories. Although direct application of the MDRD equation (or other equations) to recorded SCr measurements does not correct the measurements (primary due to changes in creatinine assays used) this does not preclude the possibility of successfully calibrating these data.

## Conclusions

This algorithm can identify laboratory changes in eGFR calculation methods. Failure to identify these changes in method may cause misclassification of CKD and misconstrue renal function changes over time. Whilst there is scope to improve clinical coding this can only be prospective, and flagging the limitations of data at the time is important for future researchers if they are to derive most meaning from these data. Researchers using longitudinal routinely collected renal function data should account for these effects.